\documentclass[aip,preprint,superscriptaddress]{revtex4-1}
\usepackage{graphicx}
\usepackage{subfigure}  
\usepackage{booktabs}
\usepackage{amssymb}
\usepackage{epstopdf}
\usepackage{lineno,hyperref}
\modulolinenumbers[5] 

\draft 

\begin{document}


\title{Optical and electronic properties of sub-surface conducting layers in diamond created by MeV B-implantation at elevated temperatures}



\author{L.H.~Willems~van~Beveren}
\email[]{laurensw@unimelb.edu.au}
\affiliation{School of Physics,  University of Melbourne, Parkville, VIC 3010, Australia}
\altaffiliation{Now at: National Measurement Institute, Lindfield, NSW 2070, Australia}
\author{R. Liu}
\affiliation{SIMS Facility, Office of the Deputy-Vice Chancellor (Research and Development)
Western Sydney University, Locked Bag 1797, Penrith, NSW 2751, Australia}
\author{H.~Bowers}
\author{K.~Ganesan}
\author{B.C.~Johnson}
\author{J.C.~McCallum}
\author{S.~Prawer}
\affiliation{School of Physics, University of Melbourne, Parkville, VIC 3010, Australia}

\hyphenation{su-per-con-duc-ting va-cu-um spec-tro-sco-py va-lu-es da-ma-ge cor-re-spon-ding sig-ni-fi-cant pro-mi-sing vol-ta-ge con-duc-ti-vi-ty mag-ne-to-re-sis-tan-ce tem-pe-ra-tu-re}

\date{\today}

\begin{abstract}
Boron implantation with in-situ dynamic annealing is used to produce highly conductive sub-surface layers in type IIa (100) diamond plates for the search of a superconducting phase transition. Here we demonstrate that high-fluence MeV ion-implantation, at elevated temperatures avoids graphitization and can be used to achieve doping densities of 6 at.\%. In order to quantify the diamond crystal damage associated with implantation Raman spectroscopy was performed, demonstrating high temperature annealing recovers the lattice. Additionally, low-temperature electronic transport measurements show evidence of charge carrier densities close to the metal-insulator-transition. After electronic characterization, secondary ion mass spectrometry was performed to map out the ion profile of the implanted plates. The analysis shows close agreement with the simulated ion-profile assuming scaling factors that take into account an average change in diamond density due to device fabrication. Finally, the data show that boron diffusion is negligible during the high temperature annealing process.
\end{abstract}


\maketitle 

\section{Introduction}
Heavily boron (B) doped diamond structures without electronically compensating effects have been theoretically predicted to become high-temperature superconductors provided the substitutional concentration is high enough~\cite{Moussa:2008}. Indeed, in the last decade, superconductivity in B-doped diamond has been observed for single crystal~\cite{Ekimov:2004,Bustarret:2004b,Takano:2004,Takano:2007,Takano:2009,Kawano:2010a} and also for nanocrystalline diamond films~\cite{Nesladek:2006}. Chemical vapour deposition (CVD) growth in the $\langle$100$\rangle$ direction is able to incorporate boron up to 10~at.\%~\cite{Takano:2007}, but a significant amount of hydrogen incorporation into the crystal compensates much of the boron and limits the superconducting transition temperature $T_{c}$ to about 6~K. Recently, signatures of high $T_{c}$ values around 25~K were reported in higher quality CVD grown heavily boron-doped single-crystal diamond~\cite{Okazaki:2014}, corresponding to a carrier concentration of $\sim$3.1$\times$10$^{21}$~cm$^{-3}$ (=1.8~at.$\%$). Interestingly, according to their calculations, if substitutional doped boron could be arranged periodically or the degree of disorder reduced, a $T_{c}$ of approximately 100~K could be achieved via minimal percent doping.

Previous efforts to find a superconducting phase transition in boron-doped diamond through ion-implantation were unsuccessful, c.f. the work of Heera et al.~\cite{Heera:2008}. However, it is worthwile revisiting this methodology with new insights. In particular, ion-implantation at MeV, in contrast to keV energies, creates a buried B doped layer with a relatively undamaged diamond capping which acts to pressurise the implanted zone and inhibits the transformation to graphite. In addition, MeV implantation has the additional benefit that the vacancy rich region is offset from the B peak, which results in a separation between the conducting layer and the electronically compensating effects.

In the present work, the possibility of creating a high-temperature superconductor via B-implanting diamond films was explored. Ion-implantation of diamond plates (through hard masks) allows complete control of doping type, concentration and depth profile, which is beneficial for device fabrication purposes. When B is introduced by high pressure high temperature (HPHT) methods, the maximum concentration is determined by the solubility of B in carbon. This limitation is overcome when using either CVD or ion-implantation. For the ion-implantation technique the fluences required usually result in graphitization~\cite{Kalish:1995,Fairchild:2012}, a process that cannot be reversed~\cite{DresselhausKalish:1992}. Here, we take advantage of (i) high energy implantation to bury the B layer deep inside the diamond where it is subjected to high internal pressures, keeping the lattice intact even under extreme implantation conditions, (ii) high temperature implantation to promote dynamic annealing during the implantation to inhibit graphitization~\cite{Lee:1978,Tsubouchi:2006a,Tsubouchi:2006b,Tsubouchi:2009}. Residual damage can then be removed by a post implantation annealing step~\cite{Kalish:1999}. The result we report herein is a very high concentration B doped layer in a largely defect free single crystal diamond layer. We report the lattice structure and electrical properties of B doped samples fabricated by high temperature high energy high-fluence B implantation followed by high temperature annealing (HTA).

\section{Fabrication process}
2 MeV boron ions were implanted into 2 diamond plates held at a temperature of 600$^{\circ}$C: The first plate (A) was implanted to 2$\times$10$^{16}$~B~cm$^{-2}$ and a second plate (B) to 1$\times$10$^{17}$~B~cm$^{-2}$. The fluence for these implants should in theory allow the charge density to reach the metal-insulator-transition (MIT) at about 4.5$\times$10$^{20}$~cm$^{-3}$ (=0.26~at.$\%$)~\cite{Tshepe:2004,Klein:2007}, assuming a high level of activation. For CVD grown films a superconducting phase transition was observed for a critical boron density~\cite{Kawano:2010b} of 3$\times$10$^{20}$~cm$^{-3}$. Post-implant, these plates were Bristol acid boiled, to remove the surface graphitisation~\footnote{The Raman signal from graphite can easily dominate the signal associated with implantation damage.}, and annealed in vacuum at a temperature of $T$=1300$^{\circ}$C for 10-15~minutes to activate the boron ions and recover the diamond lattice from implantation damage. To create robust, low-resistance electrical contacts to the sub-surface layer a laser milling and back brazing process was developed~\cite{Lichter:2015}. The diamond plates were patterned with 4 electrodes (with a 500~$\mu$m contact spacing) to form a van der Pauw square~\cite{Pauw:1958} device configuration. To obtain an estimate for the thickness of the buried layer we relied on the doping profile (full width half maximum) from Stopping-and-Range-of-Ions-in-Matter (SRIM) simulations~\cite{Zeigler:1985}, predicting a sheet thickness of $t_{s}$=$\sim$100~nm~\footnote{The SRIM simulation predicts (Lortentzian fit) a full-width-half-maximum of the doping peak of $\sim$75~nm. If we include longitudinal straggle, $\Delta R_{p}$=$\sim$25~nm, we obtain an estimated sheet thickness of $\sim$100~nm.} at a depth of $R_{p}$=1.37~$\mu$m below the diamond surface (projected range). Since all implants are the same energy, the sheet thickness becomes a scaling factor that affects all measurements equally. Importantly, the SRIM simulation predicts a peak B concentration of 2$\times$10$^{21}$~B~cm$^{-3}$, and 1$\times$10$^{22}$~B~cm$^{-3}$, for a fluence of 2$\times$10$^{16}$~B~cm$^{-2}$ and 1$\times$10$^{17}$~B~cm$^{-2}$, respectively. These concentrations correspond to an atomic boron percentage of 1.2~at.$\%$, and 6~at.$\%$, respectively. This result will later be used for calculating the boron activation fraction.

\section{Experimental results}
\subsection{Secondary ion mass spectrometry}
To map out the doping profile as a function of depth (after device fabrication and electronic characterization), dynamic secondary ion mass spectrometry (SIMS) was performed on plates A and B. For this analysis, in a Cameca IMS 5fE7 instrument, a primary ion oxygen ($O^{+}_{2}$) beam was used with an impact energy of 7.5~keV and a beam current of 200~nA to raster a 180~$\mu$m x 180~$\mu$m region of the surface. Using these settings, a total depth of 2.27~$\mu$m was profiled with an analysis area of 33~$\mu$m in diameter. The diamond plates were gold coated beforehand and an in-situ electron beam was used to prevent sample charging during the analysis. The secondary ions included $^{11}$B$^{+}$, $^{12}$C$^{+}$, $^{11}$B$^{12}$C$^{+}$, and $^{12}$C$_{2}$$^{+}$. The  SIMS analysis was always undertaken centrally on the sample area that was characterized electrically. The sputter rate was determined by assessing the depth of the analysis crater using a KLA Tencor Alpha-Step IQ profilometer and was determined to be 0.474~nm/s.

Here, we used a 'standard' sample with a known boron concentration to work out the relative sensitivity factor (RSF). We also used a second method to obtain the RSF by integrating the SIMS spectrum to yield the value for the fluence used for that implant. Even though the first method was used to relate the SIMS intensity to a B concentration, the ion profiles produced by the two methods were very similar. The resulting spectra were then compared to the SRIM predicted ion profile for plate B (black), c.f. Fig.~\ref{SIMS_SRIM}. The SIMS analysis shows that the peak concentration for plate B (blue) is close to 1$\times$10$^{22}$~B~cm$^{-3}$, in agreement with SRIM simulations. The difference $dz$ in projected range $R_{p}$ can be corrected for when taking into account a scaling factor, which reflects a reduction in diamond density resulting from the device fabrication. The scaling factors are 0.89 for plate A and 0.847 for plate B, respectively. The effective diamond density in the cap can be calculated by multiplying the scaling factor with the diamond density (3.52~gr cm$^{-3}$). The peak concentration ratio between plate B and A (red) is 3.7x, whereas it was expected to be a factor of 5x (The integrated areas under the SIMS peaks scale as 3.9x.). This difference can be explained by the error in setting the fluence for each of the separate timed implantation runs. The full-width-half-maximum (FWHM) of the boron peak concentration for both plates yields a thickness of $\sim$90~nm, close to the predicted 100~nm, which demonstrates limited diffusion of boron during high temperature annealing.

\subsection{Optical spectroscopy}
Room temperature (RT) Raman and photoluminescence (PL) spectroscopy were performed to determine the crystal damage after (hot) implantation~\cite{Hohne:2007} and after annealing, respectively. The excitation laser used had a wavelength of 532~nm.

Figures~\ref{PL_before_after}(a)-(b) show the PL spectra at each stage of the processing, i.e. pristine, hot implanted and high-temperature annealed. These spectra are normalized to the diamond peak intensity and offset for clarity. The main feature is the sharp peak at 572~nm, corresponding to the first order Raman line, and the photoluminescence signals related to nitrogen vacancy centers (NV$^{-}$ at 637~nm and NV$^{0}$ at 575~nm) in the crystal. The NV peaks are clearly visible before and after the implantation and annealing steps.

To investigate the implantation damage and effect of annealing, Raman spectra were taken, as shown in Figs.~\ref{Raman_before_after}(a)-(b). For plate A, c.f. Fig.~\ref{Raman_before_after}(a) (red) there are three peaks visible in the 1400-1800~cm$^{-1}$ range, resulting from the ion-implantation process: There is a small peak visible near 1451~cm$^{-1}$ (inset) and a larger peak at around 1501~cm$^{-1}$, which appear to be unique to high-energy implantation into diamond~\cite{Hunn:1995}. A third large peak at 1637~cm$^{-1}$ has previously been attributed to a split-interstitial~\cite{Orwa:2000,Walker:1997} and anneals out above 1300$^{\circ}$C. Interestingly, the implantation process quenches the NV$^{0}$ PL signal (observed at 1427~cm$^{-1}$). No significant shift of the diamond peak was observed.

In the Raman spectrum of plate B, c.f. Fig.~\ref{Raman_before_after}(b) (red), we again observe several peaks at 1498~cm$^{-1}$ and 1637~cm$^{-1}$ consistent with the presence of vacancies and split interstitials~\cite{Prawer:2004,Prawer:2004b}. The Raman band around $\sim$500~cm$^{-1}$ was previously assigned to boron dimers B$_{2}$ and to clustered boron atoms~\cite{Szirmai:2012}. Post HTA, there are no more defect peaks visible in the 1000-2000~cm$^{-1}$ waveshift range for plate A, except for a 1559~cm$^{-1}$ peak (red trace), which is usually attributed to disordered sp$^{2}$ bonded carbon atoms. The Raman spectrum for plate B post-HTA indicates that the high temperature annealing repairs most of the damage associated with the ion-implantation process as no more damage peaks can be observed. Furthermore, post HTA, the diamond peak, as well as the NV$^{0}$ PL peak increase in intensity (when the data is not normalized to the diamond peak). We therefore conclude that the diamond lattice has been preserved as far as Raman spectroscopy is concerned. In particular, despite the extremely high fluence of 1$\times$10$^{17}$~B~cm$^{-2}$ there is no evidence of graphitization. The hot MeV implantation followed by a HTA method allows very high concentrations of B to be introduced into the lattice (see Table~\ref{tab:data_summary}), while still maintaining the integrity of the diamond lattice structure.

\subsection{Electronic transport measurements}
Low temperature electronic transport measurements were performed as a function of temperature and magnetic field. From these measurements we extract thermally activated behaviour of the conductivity. Furthermore, from Hall measurements we extract a charge density and carrier mobility, which can be compared to values extracted from magneto-resistance. To this end we used a cryogen-free dilution refrigerator (Leiden Cryogenics) with insertable probe to quickly cooldown samples from RT down to $\sim$4-5~K. The system incorporates a superconducting magnet for both magneto-resistance and Hall measurements. For RT measurements we used a permanent magnet. The methodology for calculating the `bulk' resistivity $\rho_{xx}$ of the sub-surface conducting layer was as follows. At room temperature, current-voltage (I-V) measurements were taken to extract the 4-terminal square resistance ($R_{4T}=V_{xx}/I_{sd}$) of the plates, which multiplied by the buried layer thickness directly provides $\rho_{xx}$=$R_{4T}\times$ $t_{s}$.

The result of a Hall measurement on plate A is shown in Fig.~\ref{plateAvsT}(a). To obtain those data, we ramp the source-drain current $I_{sd}$ from -2 to +2~mA and record the source-drain voltage $V_{sd}$ and the Hall voltage $V_{xy}$ simultaneously for both +0.34~T and -0.34~T. The data corresponding to opposite magnetic fields are then subtracted (symmetrized) and divided by 2 to obtain the effective Hall voltage $\Delta V_{xy}$=0.5($V_{xy,+B}$ - $V_{xy,-B}$). The resulting slope then represents the Hall resistance over a magnetic field range of $B$=0.34~T. Here, the Hall scattering factors are assumed to be unity (see Ref.~\cite{Lander:2000} and citations therein). The extracted Hall slope in units of $\Omega$/T equals 47.8~m$\Omega$/T, which corresponds to a 2D carrier density of $n_{2D}$=1.30$\times$10$^{16}$~cm$^{-2}$. Assuming a thickness of 100~nm, and with a sheet resistance of 270~$\Omega$, the 3D density equals $n_{3D}$=(1.30~$\pm$~0.008) $\times$10$^{21}$~cm$^{-3}$ and the resistivity equals $\rho_{xx}$=2.7~m$\Omega$-cm with a mobility of $\mu$=(1.77~$\pm$~0.01)~cm$^{2}$V$^{-1}$s$^{-1}$. This carrier density is very close to the SRIM prediction that for this fluence the peak B-concentration is $n_{3D}$$\sim$2x10$^{21}$~cm$^{-3}$. This would indicate that we have a reasonably good (thermal) activation of boron ions at RT. The fraction of active B can be calculated directly based on the predicted peak B-doping concentration and the RT Hall data to be 1.30$\times$10$^{21}$ / 2$\times$10$^{21}$ = 65$\%$.

Figure~\ref{plateAvsT}(b) shows $ln(\sigma_{xx}$) as a function of $1000/T$ between 100 and 25~K. From this Arrhenius plot we extract an activation energy of $E_{A}$=(0.86~$\pm$~0.1)~meV. This value is much less than $E_{A}$=0.37~eV~\cite{Thonke:2003}, which is typically cited for B-doped diamond. This inconsistency does not necessarily imply that the conduction observed here is due to some defect other than B. In~\cite{Borst:1996gb} it was shown that for very low resistance samples the exponential dependence on inverse temperature shifts to higher temperatures and the corresponding ionization energy becomes smaller and smaller in accordance with a metal-to-insulator transition. The inset shows how the conductivity decreases with lowering the temperature on a linear scale. The two-terminal resistance of plate A ($R_{2T}=V_{sd}/I_{sd}$) as a function of temperature in the Kelvin regime is shown in Fig.~\ref{plateAvsT}(c). The sample shows a significant resistance increase below 4~K. We expect that upon cooling down to the mK regime, charge carriers localize in electronic trap states, resulting in a freeze out that increases the resistance exponentially. In fact, plate A becomes so resistive below $\sim$2~K, that it is not easily measurable by our electronic setup. As a consequence, c.f. Fig.~\ref{plateAvsT}(d), the sample resistivity $\rho_{xx}$ goes into compliance (dashed line) for $T<7$~K. This behaviour corresponds to an insulating material. 

Next, the electrical characterization of plate B is presented. The 2-terminal resistance at room temperature is $R_{2T}$=529~$\Omega$ and the 4-terminal resistance $R_{4T}$=47~$\Omega$. This tells us that the contact resistance is significant here and the conductivity of the material itself is low. Since these values are much lower than those for plate A, it confirms our expectations for implantation to a higher fluence. The RT carrier density of plate B was measured again by symmetrization. Since plate B is likely to have a higher carrier density it is harder to extract the Hall voltage (since the Hall coefficient is inversely proportional to the carrier concentration), especially with small values of magnetic field available. However, the carrier density was estimated by ramping the current over a 10~mA range (not shown). The room temperature resistance of R$_{4T}$=47~$\Omega$ corresponds to a resistivity of $\rho_{xx}$=$t_{S}$$R_{4T}$=0.47~m$\Omega$-cm and yields a mobility of $\mu$=1/(e $\rho_{xx}$ $n_{3D}$)=(3.89~$\pm$~0.5)~cm$^{2}$V$^{-1}$s$^{-1}$, respectively, assuming a sheet thickness of $t_{S}$=100~nm and with a RT carrier density of $n_{3D}$=(3.37~$\pm$~0.5)$\times$10$^{21}$~cm$^{-3}$. The RT carrier density extracted is not too different from plate A, which could indicate thermal activation of carriers. The fraction of active B for plate B at RT can once again be calculated based on the peak B-doping concentration and the RT Hall data to be 3.37$\times$10$^{21}$ / 1$\times$10$^{22}$ = 33.7$\%$, which indicates significant activation, but less than the fraction observed for plate A.

An Arrhenius plot of the conductivity for plate B is shown in Fig.~\ref{plateBvsT}(a). The extracted activation energy is $E_{A}$=(1.5~$\pm$~0.1)~meV, which is again significantly lower than 0.37~eV~\cite{Thonke:2003}, which agrees~\cite{Borst:1996gb} with the fact that  our measured carrier density is much larger than $\sim$10$^{19}$~cm$^{-3}$. We then cooled plate B further down and mapped out $\rho_{xx}$ as a function of temperature c.f. Fig.~\ref{plateBvsT}(b). The increase in resistivity with decreasing temperature is much smaller (note that we plot m$\Omega$-cm here) than for plate A, which was shown in Fig.~\ref{plateAvsT}(d).

We used the symmetrization technique, explained earlier, to extract Hall data at low temperatures~\footnote{Temperature changes caused by sweeping the magnetic field make it difficult to accurately extract values for carrier density and mobility.}. Here, we now set our superconducting magnet to a fixed field of $B$=1~T and ramp a current through the sample from $I$=-0.5 to +0.5~mA. The extracted Hall slope (not shown) of 0.991~$\Omega$/T corresponds to a $n_{2D}$=6.30$\times$10$^{14}$~cm$^{-2}$. Assuming 100~nm thickness for the buried layer we find $n_{3D}$=(6.30~$\pm$~0.43)$\times$10$^{19}$~cm$^{-3}$ for the carrier density at 4-5~K. The corresponding resistivity and mobility are $\rho_{xx}$=1.71~m$\Omega$-cm and $\mu$=(57.9~$\pm$~3.7)~cm$^{2}$V$^{-1}$s$^{-1}$, respectively.

Comparing the carrier density at RT vs 4-5~K it appears as the cooldown reduces the density from $n_{3D}$=$\sim$1$\times$10$^{21}$~cm$^{-3}$ to $\sim$1$\times$10$^{19}$~cm$^{-3}$, which is about 2 orders of magnitude difference, which could explain the corresponding mobility increase. We then cooled down plate B from 6~K to base temperature while recording the sample resistivity. From Fig.~\ref{mKdata}(a) there does not appear to be any evidence of a superconducting transition, i.e. where the resistivity would drop to zero below a critical temperature.

However, unlike plate A the sample resistance does not increase exponentially to values $\gg$~k$\Omega$, which may imply that we are close to the metal-insulator transition. Measurements taken at 120~mK c.f. Figs.~\ref{mKdata}(b)-(c) demonstrate the I-V traces at $B_{z}$=0~T remain linear and do not show a superconducting gap, even under low-biasing conditions~\cite{Nesladek:2006}. From the low-temperature I-V traces it is evident that the contact resistance dominates the total resistance (by about a factor of 10) and not the intrinsic resistance of the buried layer itself. This was also observed in the data taken at RT. The mK resistivity of plate B equals $\sim$2.29~m$\Omega$-cm, which is roughly 5 times its RT value.

We also performed magnetoresistance (MR) measurements of plate B, as shown in Fig.~\ref{mKdata}(d). The data show a positive MR (PMR)~\cite{Russell_Leivo:1972,Wang:2000,Bustarret:2004a} with a quadratic dependence on the applied magnetic field~\cite{Wedepohl:1957,Bate:1959}. However, note that the effect is small: over a field range of 6~T the change in resistivity is only $\sim$0.2 m$\Omega$-cm. This type of dependence (PMR) suggests a localisation of the wave functions on boron atoms. Basically, with increasing magnetic field, a shrinkage of the wave functions leads to a progressive localisation resulting in decrease of the conductivity, i.e. a PMR~\cite{Chambers:1990}. Mathematically, the quadratic dependence arises from the ($\omega_{c}\tau$)$^{2}$=($\mu$$B_{z}$)$^{2}$ term in the conductivity tensor~\cite{Beenakker:1991}. Note that this behaviour is quite different than the negative MR (NMR) usually observed for other carbon-based systems such as N-doped ultra-nanocrystalline diamond (N-UNCD) films~\cite{Nesladek:2006b} ($n$-type in 3D) or hydrogen terminated diamond surfaces ($p$-type in 2D), where the transport is dominated by weak (anti) localization (WL) effects~\cite{Edmonds:2015}. For doping densities close to the MIT WL is less pronounced. From the parabolic MR data, we extract a carrier mobility value of $\mu$=(7.06~$\pm$~0.33)~cm$^{2}$V$^{-1}$s$^{-1}$. This value is much lower than the mobility extracted previously from Hall measurements, which could possibly be ascribed to the much larger current used, leading to self-heating and lowering of the mobility.

\section{Summary and Conclusions}
Using high temperature, high fluence, high energy B ion implantation, it is possible to produce buried layers of B doped diamond with a peak concentration of nearly 6~at.$\%$. Raman and PL spectroscopy show that graphitization can be prevented by the use of high temperature dynamic annealing during the MeV implantation. Post-implantation annealing at even higher temperatures results in removal of remaining point defects. The low temperature electrical measurements of the heavily doped samples reveal an activation energy of about 1~meV over the range from RT to about 100~K. Magnetoresistance measurements at low temperatures show evidence for localization of the electronic wavefunction (holes) on the boron atoms. Despite the very high B concentration and apparent integrity of the diamond lattice, no superconducting transition could be observed. These results suggest that higher fluences and/or more effective (in-situ) annealing techniques may be required to further eliminate electronic defects in order to realise implantation-based diamond superconductivity. Furthermore, electronic grade diamond plates with parts per billion (ppb) rather than parts per million (ppm) nitrogen content may reduce the effects of compensation doping and scattering. Nevertheless, with standard grade diamond plates very high dopant activation was achieved with RT carrier densities above the MIT critical limit.

\section{Acknowledgements}
This material is based on research sponsored by Air Force Research Laboratory under agreement number FA2386-13-1-4055. The U.S. Government is authorized to reproduce and distribute reprints for Governmental purposes notwithstanding any copyright notation thereon. The authors acknowledge A. Stacey for discussions and access to ion-implantation and/or ion-beam analysis facilities at the ACT node of the Heavy-Ion-Accelerator Capability funded by the Australian Government under the NCRIS program. The authors furthermore acknowledge the Western Sydney University's SIMS Facility.

\section{Disclaimer}
The views and conclusions contained herein are those of the authors and should not be interpreted as necessarily representing the official policies or endorsements, either expressed or implied, of Air Force Research Laboratory or the U.S. Government.



\newcommand{\noopsort}[1]{} \providecommand{\noopsort}[1]{}

\clearpage      
\begin{center}
{\large FIGURE CAPTIONS}
\end{center}
\vspace*{1cm}

\begin{figure}[htbp]
\caption{\label{SIMS_SRIM} (Color online) Secondary ion mass spectrometry of diamond plates A (blue) and B (red), together with the 2 MeV SRIM simulation of plate B (black).}
\end{figure}

\begin{figure}[htbp]
\caption{\label{PL_before_after} (Color online) Photoluminescence spectroscopy of diamond implanted with 2$\times$10$^{16}$~B~cm$^{-2}$ (a) and 1$\times$10$^{17}$~B~cm$^{-2}$ (b). Spectra are shown for the pristine (blue), as implanted (red) and annealed (black) samples.}
\end{figure}

\begin{figure}[htbp]
\caption{\label{Raman_before_after} (Color online) Raman spectroscopy of diamond implanted with 2$\times$10$^{16}$~B~cm$^{-2}$ (a) and 1$\times$10$^{17}$~B~cm$^{-2}$ (b). Spectra are shown for the pristine (blue), as implanted (red) and annealed (black) samples.}
\end{figure}

\begin{figure}[htbp]
\caption{\label{plateAvsT} (Color online) Electrical characterization: (a) Hall measurement of plate A at RT. (b) Arrhenius plot of the conductivity of plate A. (c) Two-terminal resistance $R_{2T}$ and resistivity (d) of plate A as a function of temperature using $I$=0.1~$\mu$A.}
\end{figure}

\begin{figure}[htbp]
\caption{\label{plateBvsT} (Color online) Electrical characterization: (a) Arrhenius plot of the conductivity of plate B. (b)  The resistivity $\rho_{xx}$ of plate B as a function of temperature using $I$=1~$\mu$A.}
\end{figure}

\begin{figure}[htbp]
\caption{\label{mKdata} (Color online) (a)  The resistivity $\rho_{xx}$ of plate B as a function of temperature (mK range) using $I$=1~$\mu$A. (b) $I$-$V_{sd}$ and (c) $I$-$V_{xx}$ of plate B at $T$=120~mK and $B_{z}$=0~T (d) Magnetoresistance (resistivity) of plate B at $T$=160~mK, with $I$=250~$\mu$A.}
\end{figure}

\begin{table*}[htbp]
\caption{\label{tab:data_summary} Table showing all parameters of the B-implanted samples (600$^{\circ}$C), from both experiment and simulation.}
\end{table*}

\clearpage      
\begin{figure}
\centering
\includegraphics[width=0.44\linewidth,bb=0 0 273 252]{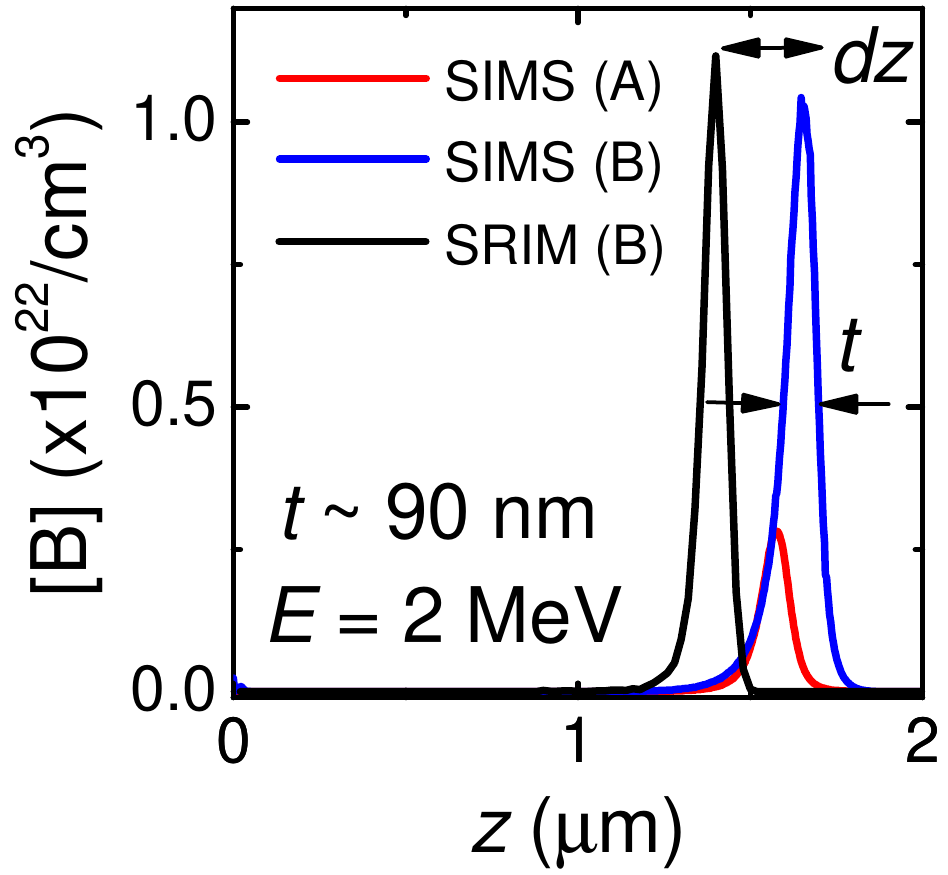} \\ 
\vspace*{1cm}
Figure 1 - L.H. Willems van Beveren, Journal Applied Physics
\end{figure}

\clearpage
\begin{figure}
\centering
\subfigure{[a]
\includegraphics[width=0.44\linewidth,bb=0 0 475 466]{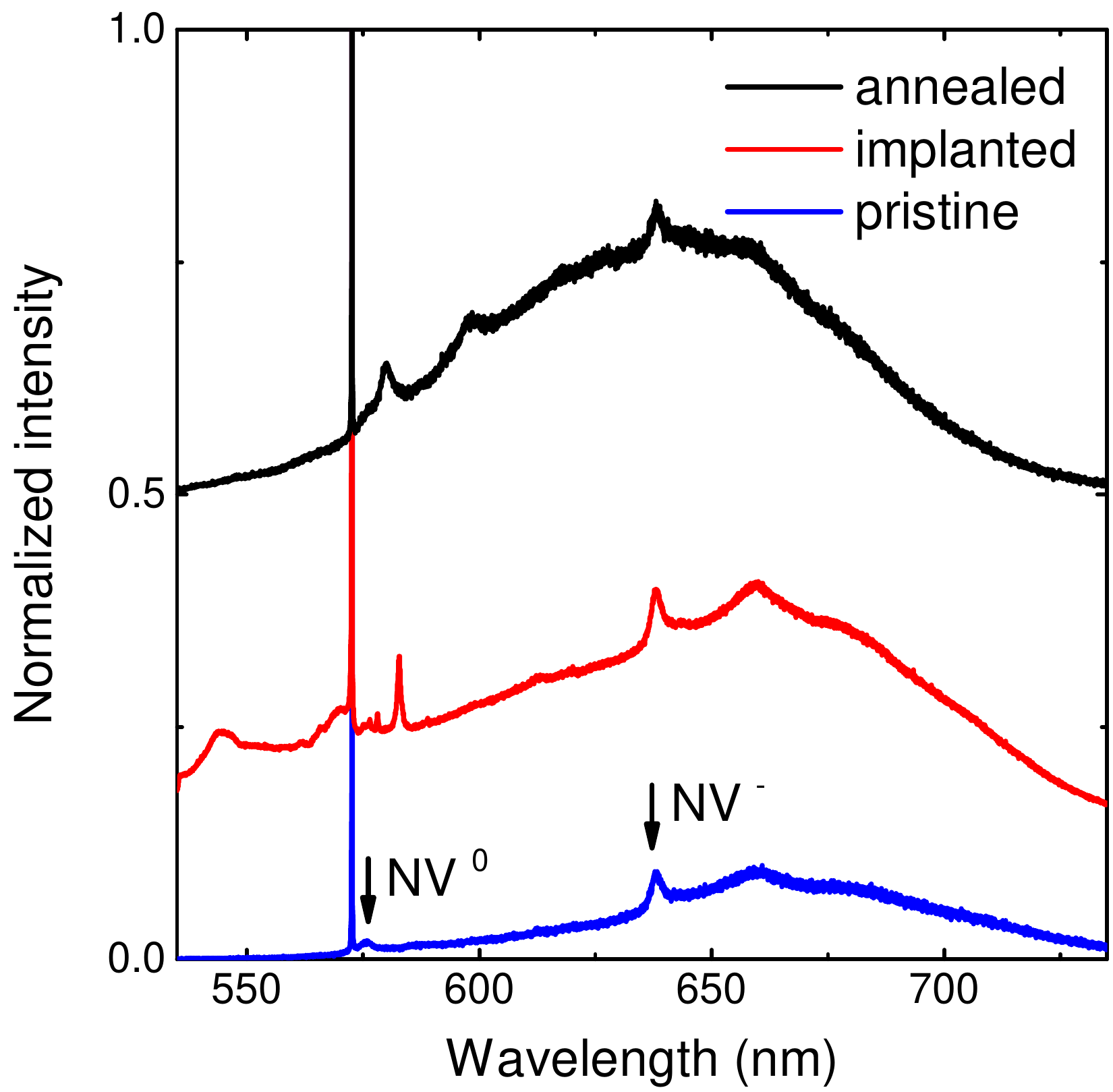}}
\subfigure{[b]
\includegraphics[width=0.44\linewidth,bb=0 0 475 466]{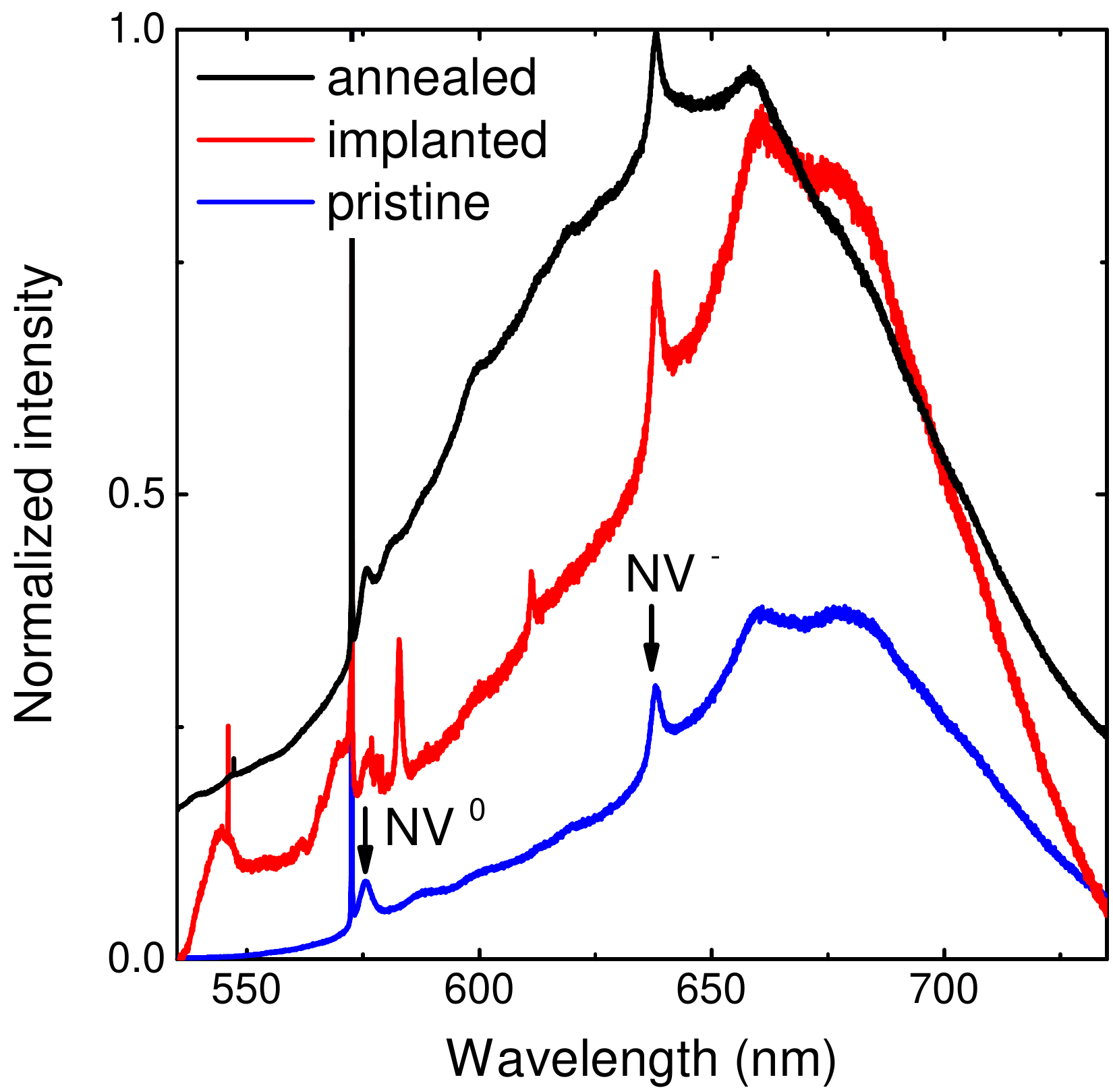}} \\
\vspace*{1cm}
Figure 2 - L.H. Willems van Beveren, Journal Applied Physics
\end{figure}

\clearpage
\begin{figure}
\centering
\subfigure{[a]
\includegraphics[width=0.44\linewidth,bb=0 0 482 725]{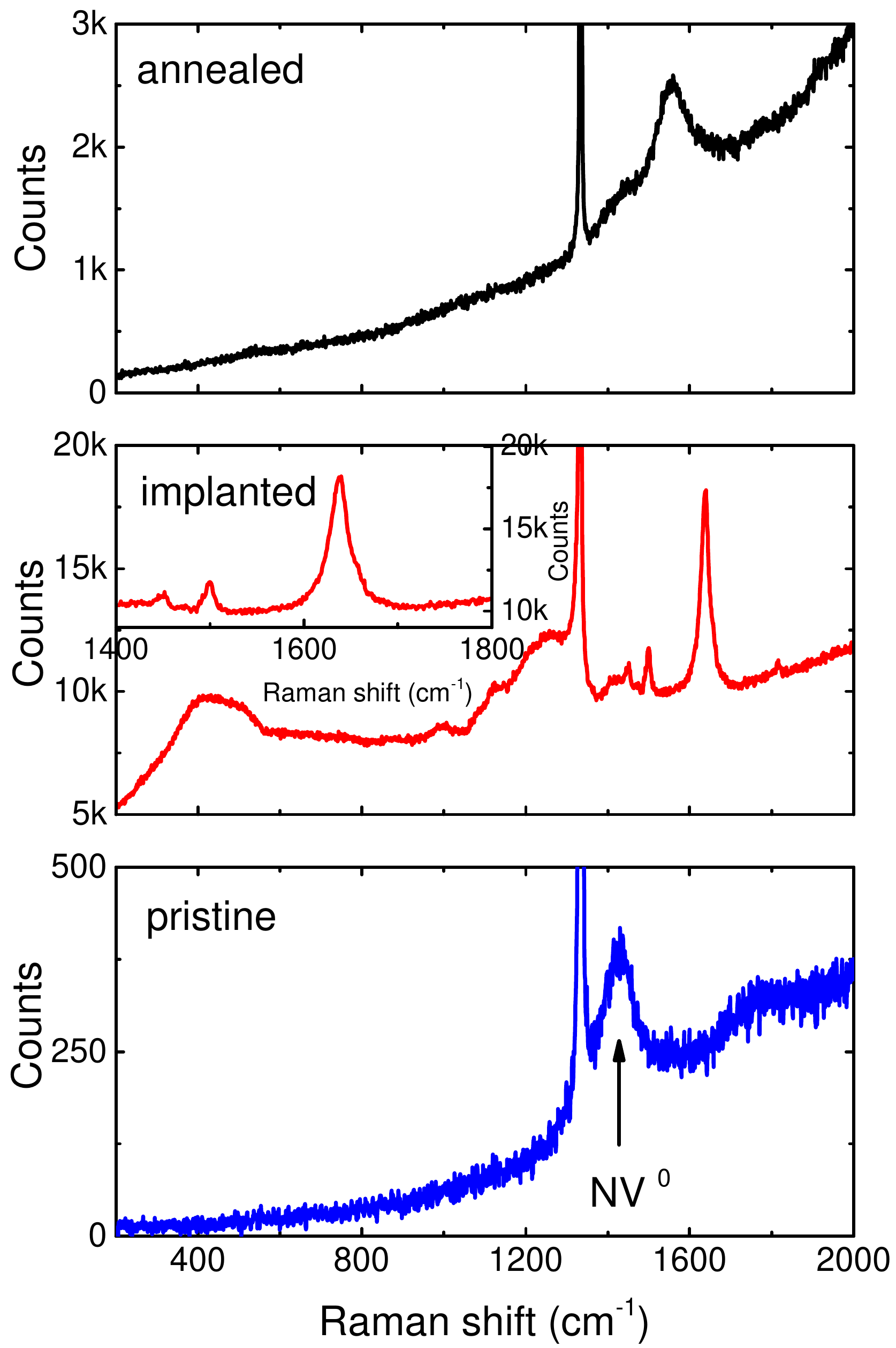}}
\subfigure{[b]
\includegraphics[width=0.44\linewidth,bb=0 0 483 725]{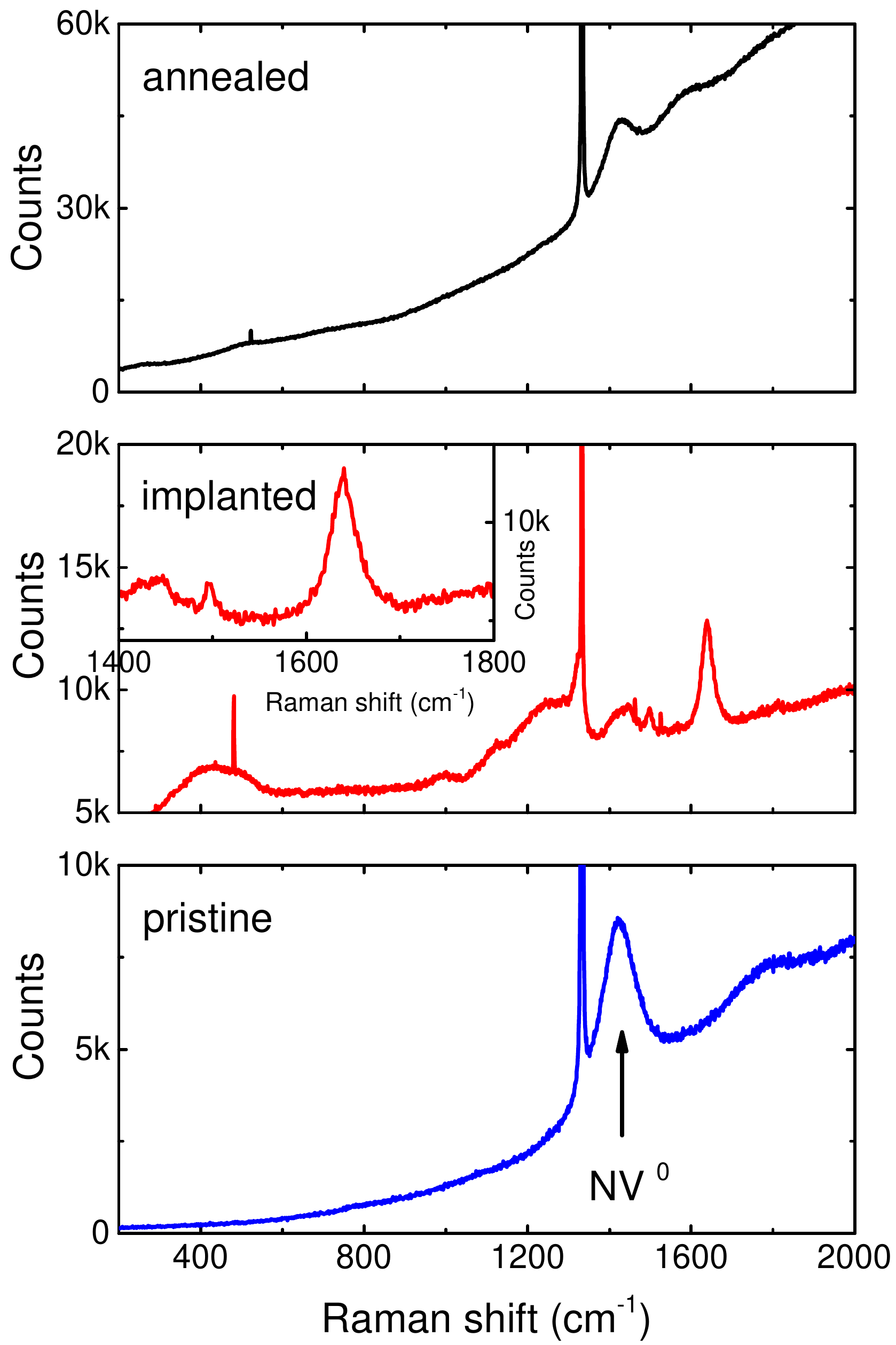}}\\
\vspace*{1cm}
Figure 3 - L.H. Willems van Beveren, Journal Applied Physics
\end{figure}

\clearpage
\begin{figure}
\centering
\subfigure{[a]
\includegraphics[width=0.44\linewidth,bb=0 0 269 268]{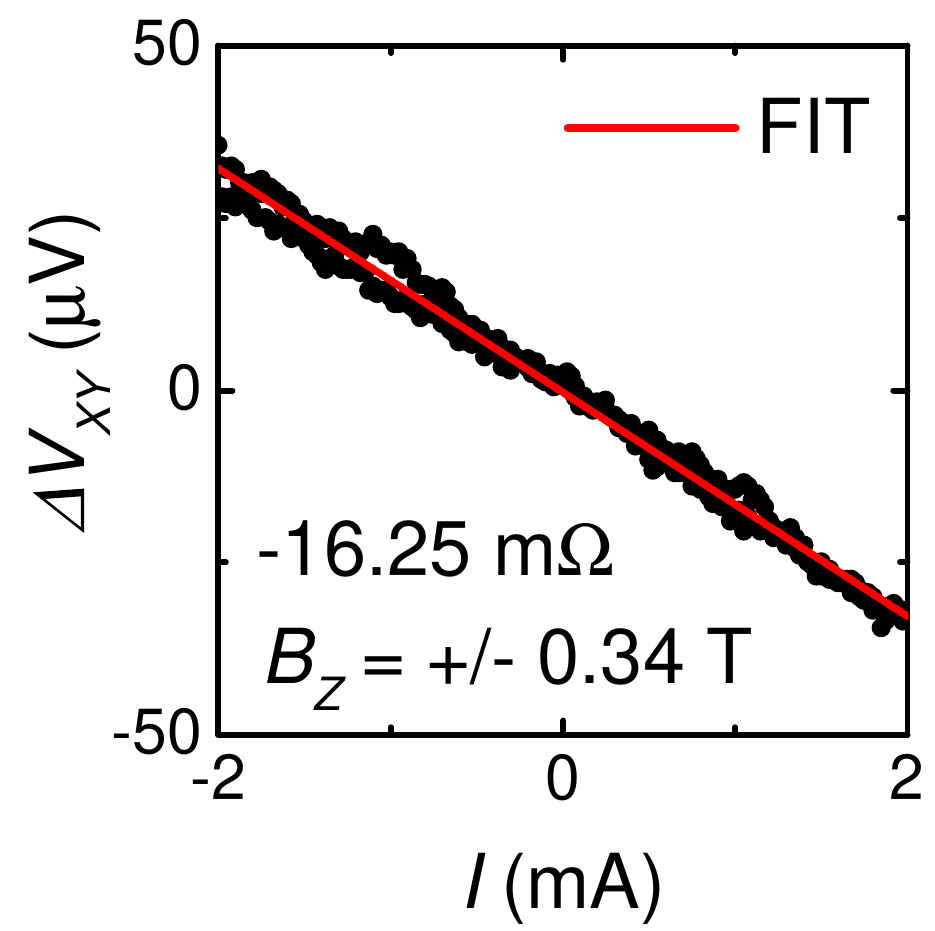}}
\subfigure{[b]
\includegraphics[width=0.44\linewidth,bb=0 0 281 291]{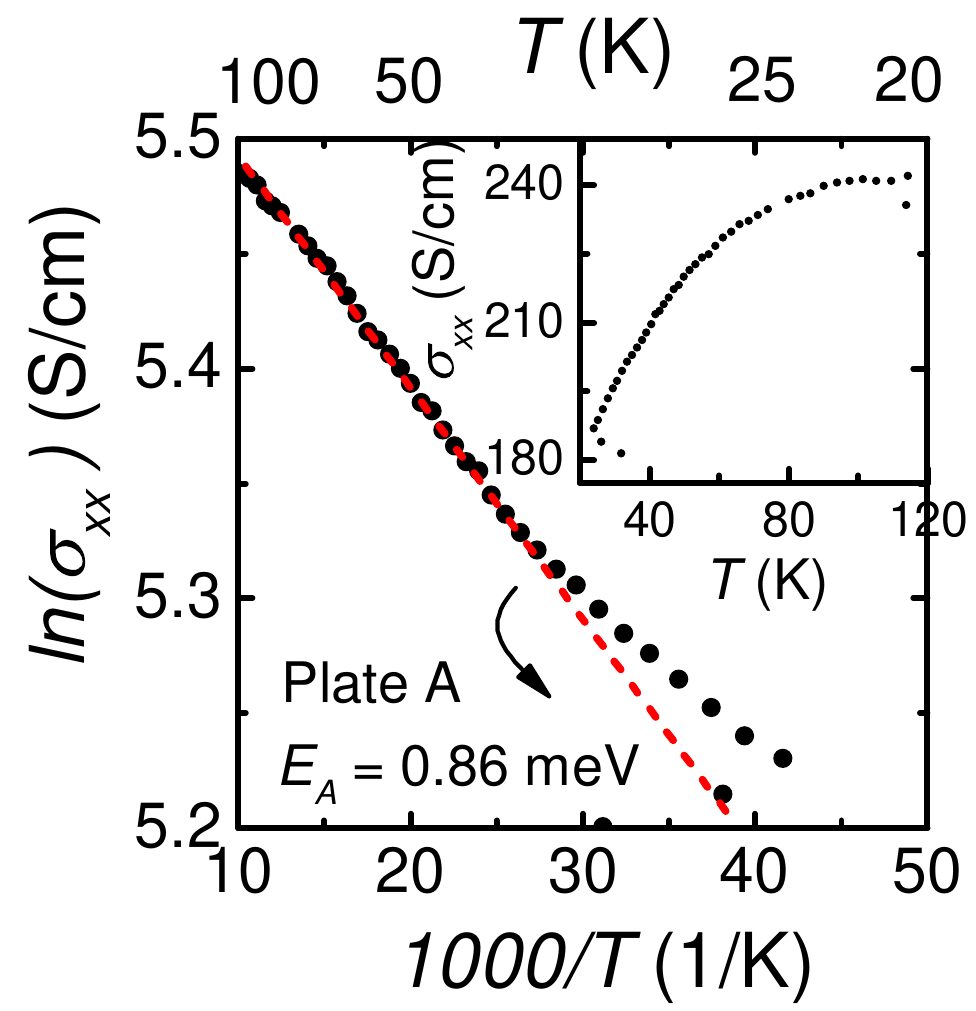}}
\subfigure{[c]
\includegraphics[width=0.44\linewidth,bb=0 0 275 268]{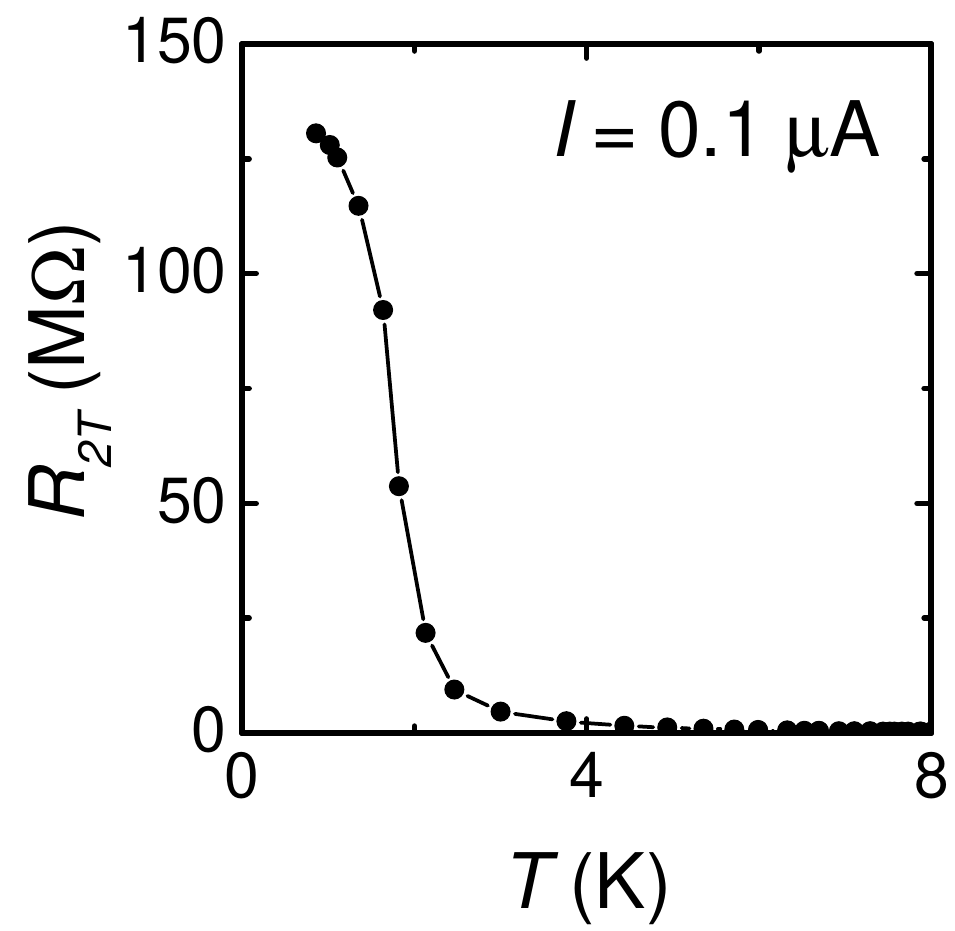}}
 \subfigure{[d]
\includegraphics[width=0.44\linewidth,bb=0 0 282 268]{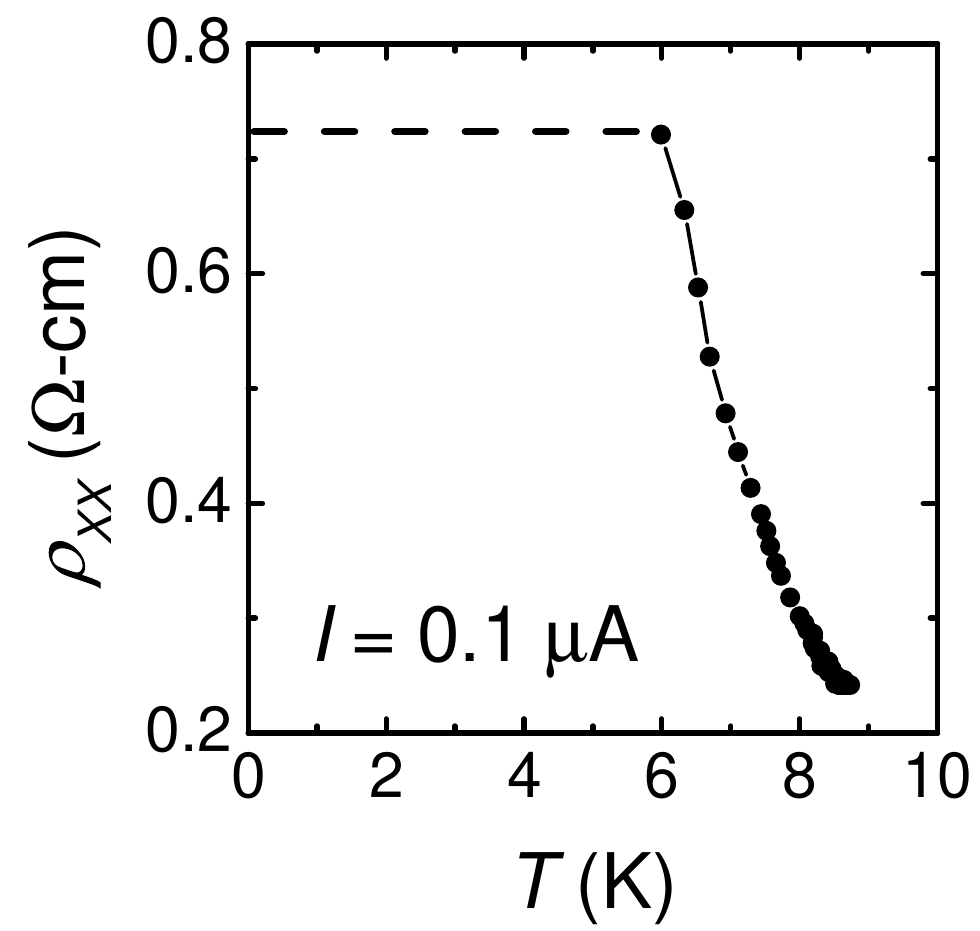}} \\

\vspace*{1cm}
Figure 4 - L.H. Willems van Beveren, Journal Applied Physics
\end{figure}

\clearpage
\begin{figure}
\centering
\subfigure{[a]
\includegraphics[width=0.44\linewidth,bb=0 0 279 289]{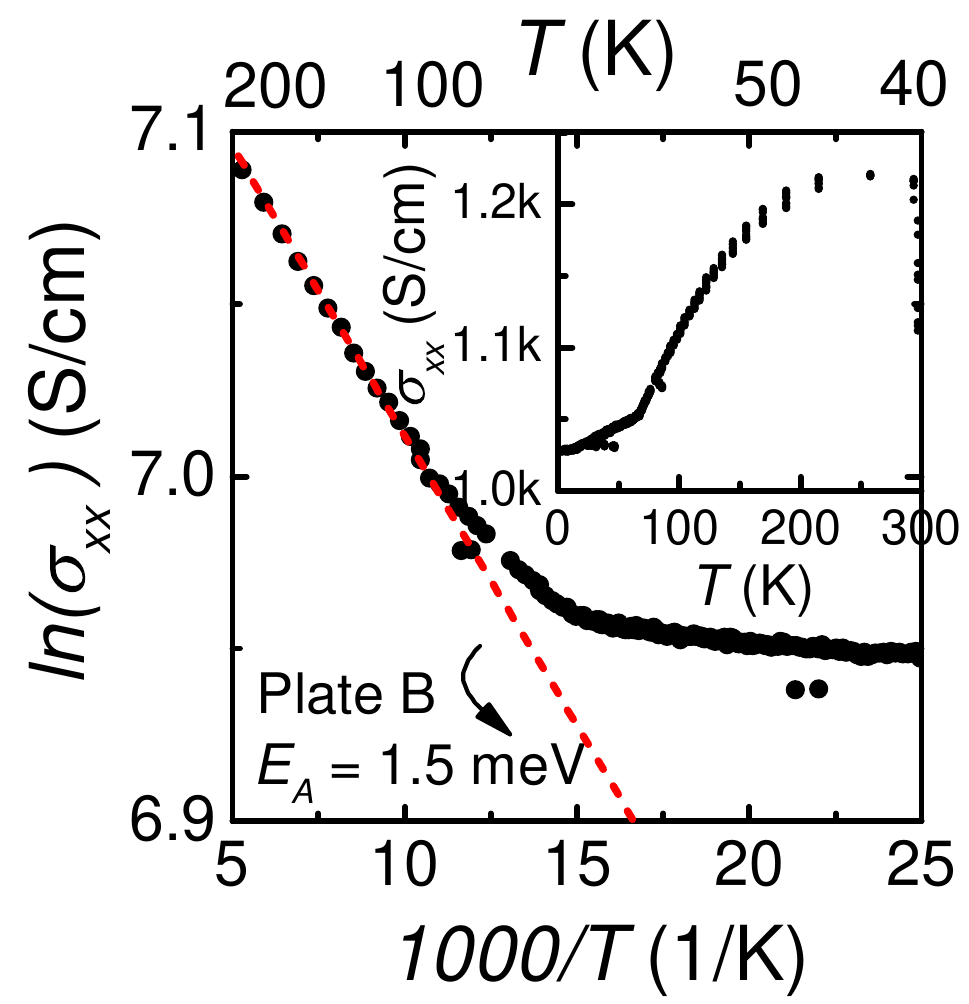}}
\subfigure{[b]
\includegraphics[width=0.44\linewidth,bb=0 0 272 257]{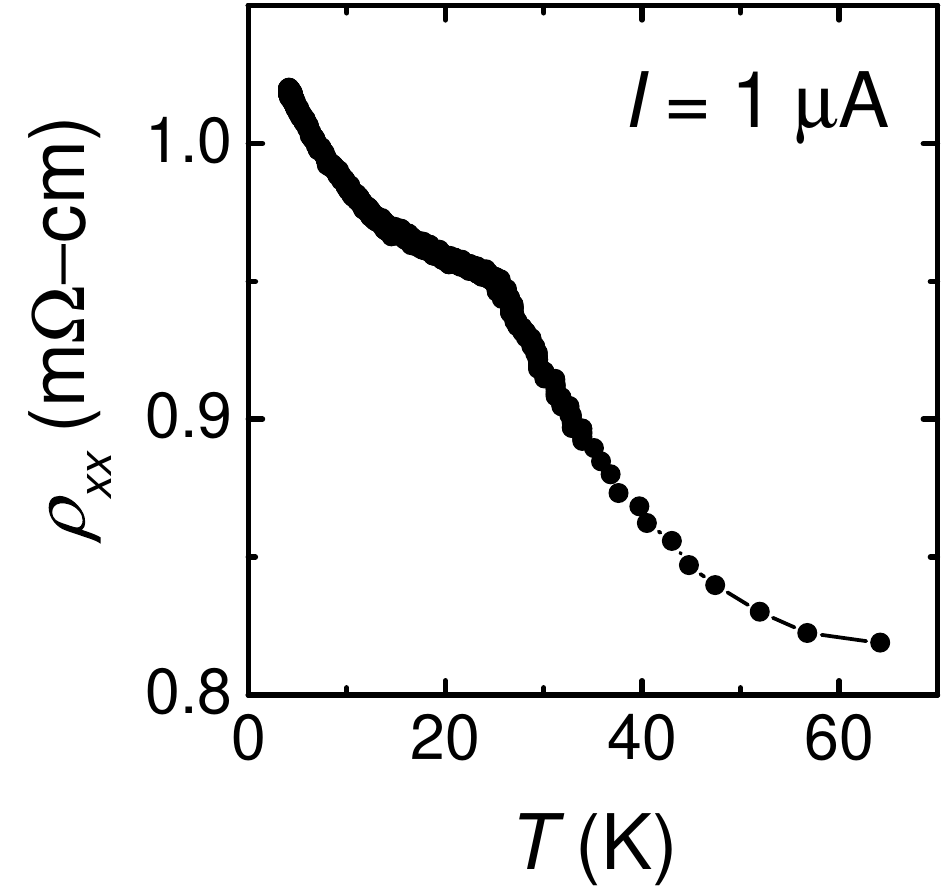}}\\
\vspace*{1cm}
Figure 5 - L.H. Willems van Beveren, Journal Applied Physics
\end{figure}

\clearpage
\begin{figure}
\centering
\subfigure{[a]

\includegraphics[width=0.4\linewidth,bb=0 0 277 257]{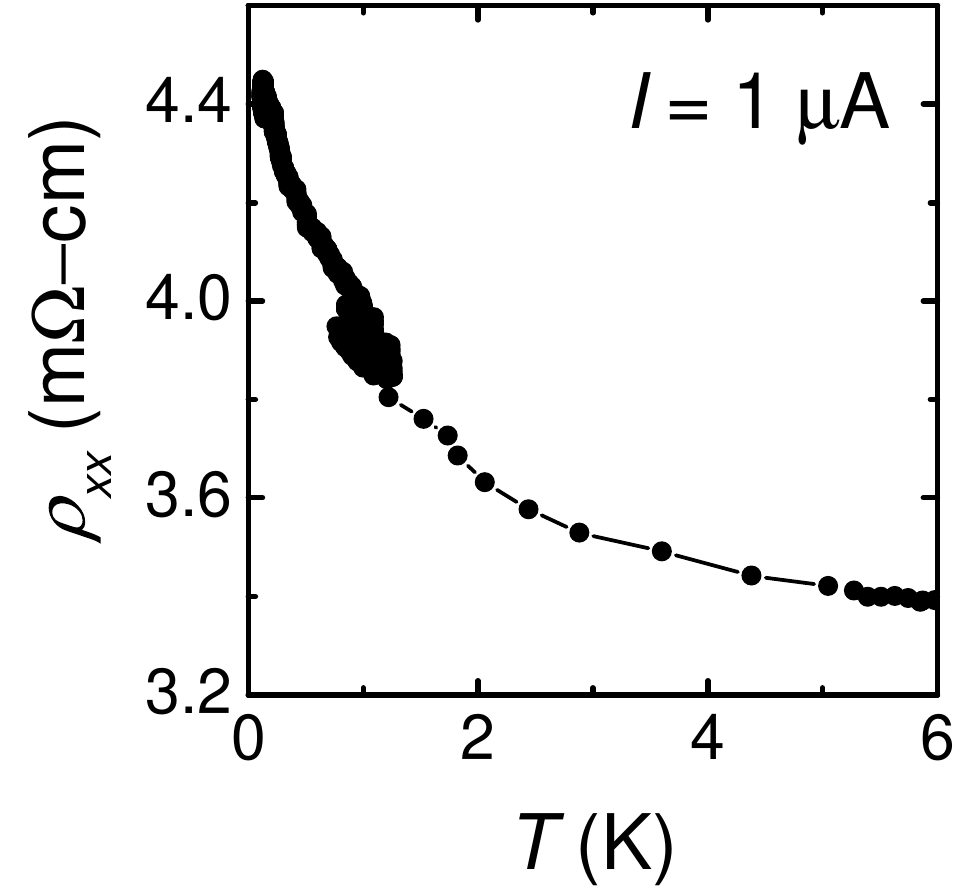}}
\subfigure{[b]
\includegraphics[width=0.4\linewidth,bb=0 0 266 259]{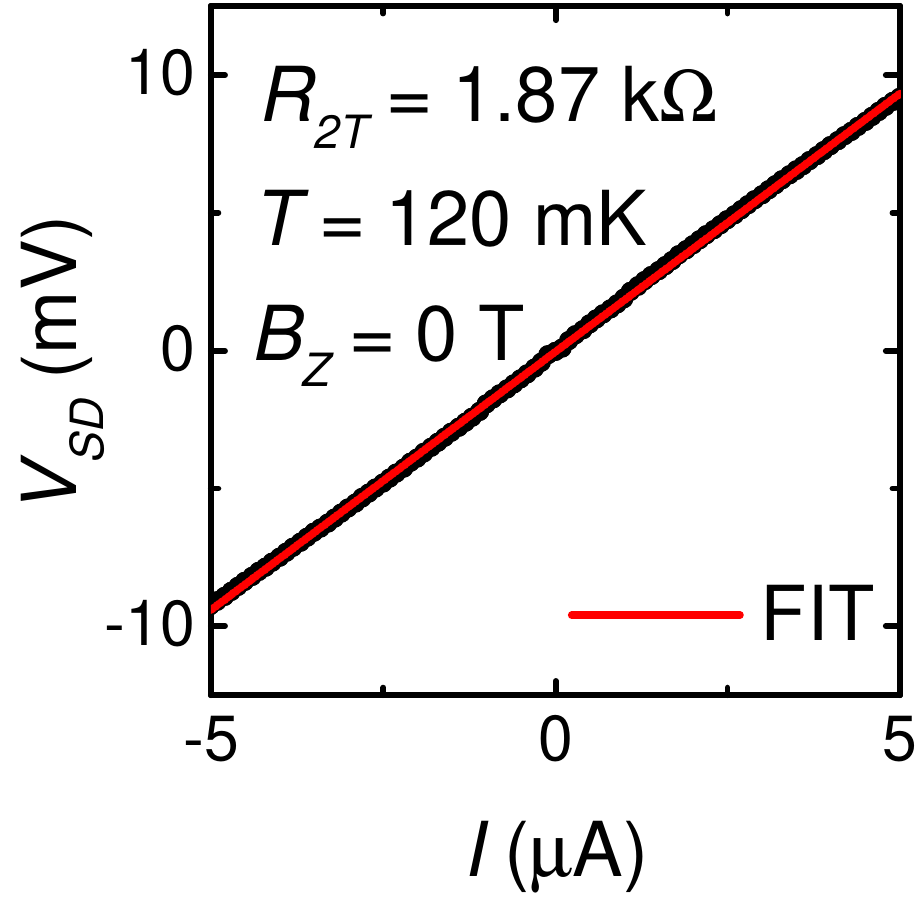}}
\subfigure{[c]
\includegraphics[width=0.4\linewidth,bb=0 0 264 259]{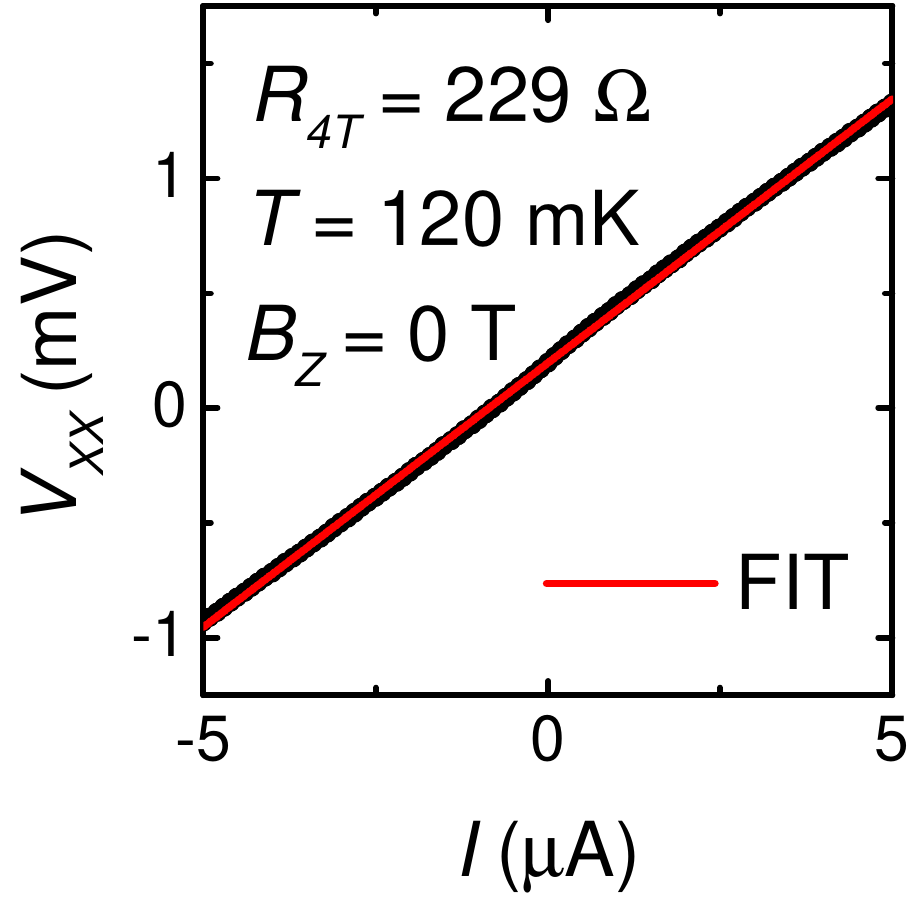}}
\subfigure{[d]
\includegraphics[width=0.44\linewidth,bb=0 0 274 266]{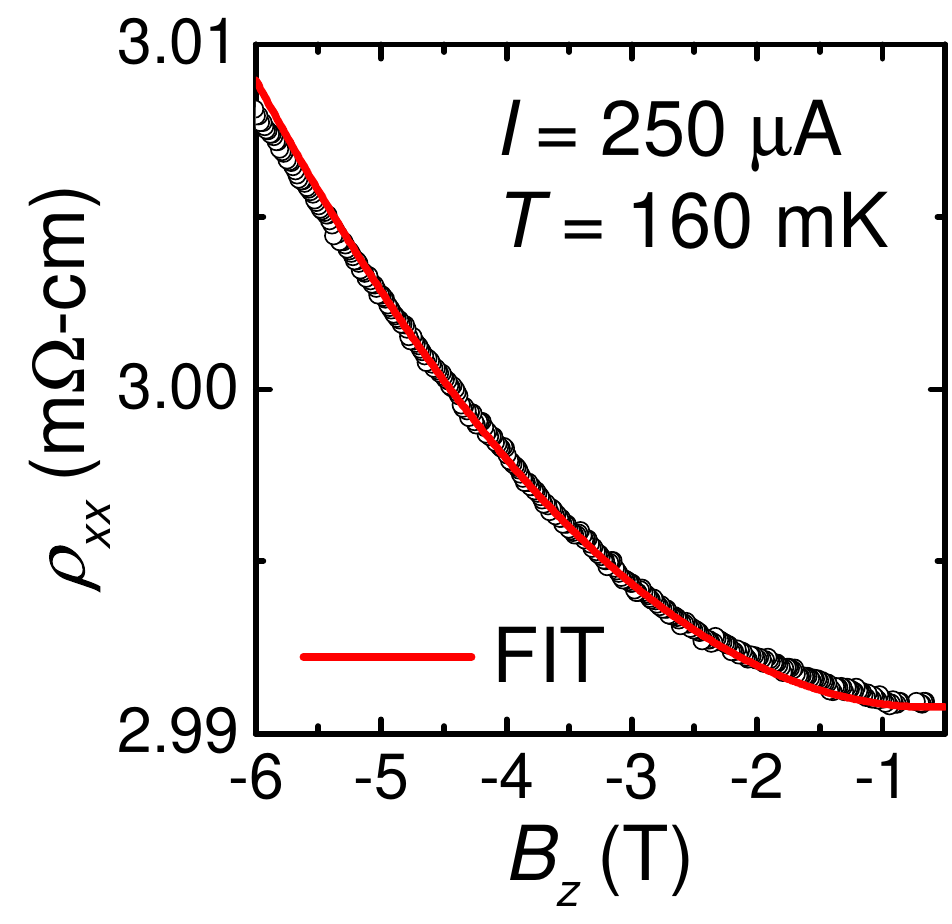}}\\
\vspace*{1cm}
Figure 6 - L.H. Willems van Beveren, Journal Applied Physics
\end{figure}

\clearpage      
\squeezetable
\begin{table*}[htbp]
\centering
 \begin{tabular*}{\textwidth}{@{\extracolsep{\stretch{1}}}*{11}{r}@{}}
  \toprule
Plate & Dose & $R_{p}$ & $\Delta R_{p}$ & [B$_{peak}$] & RT $\rho_{xx}$ & RT $n_{3D}$ & RT $\mu$ & 4K $\rho_{xx}$ & 4K $n_{3D}$ & 4K $\mu$  \\
 & [cm$^{-2}$] & [$\mu m$] & [nm] & [B~cm$^{-3}$] & [$\Omega$-cm] & [cm$^{-3}$] & [cm$^{2}$V$^{-1}$s$^{-1}$] & [$\Omega$-cm] & [cm$^{-3}$] & [cm$^{2}$V$^{-1}$s$^{-1}$] \\
  \hline
    \hline
  \midrule
A: & 2$\times$10$^{16}$ & 1.37 & 52 & 2.1$\times$10$^{21}$  & 2.70$\times$10$^{-3}$ & (1.30$\pm$0.008)$\times$10$^{21}$ & 1.77$\pm$0.01 & - & - & - \\
B: & 1$\times$10$^{17}$ & 1.37 & 52 & 1$\times$10$^{22}$  & 0.47$\times$10$^{-3}$ & (3.37$\pm$0.5)$\times$10$^{21}$ & 3.89$\pm$0.5 & 1.71$\times$10$^{-3}$ & (6.30$\pm$0.43)$\times$10$^{19}$ & 57.9$\pm$3.7  \\
  \bottomrule
\end{tabular*}
\end{table*}

\end{document}